\documentclass{amsart}%\usepackage{times}
\usepackage{amssymb}

\usepackage{graphicx,epstopdf}
\newtheorem{theorem}{Theorem}

\begin{document}
\title
{Lam\'e equation, quantum top and elliptic Bernoulli polynomials}%
\author{M-P. Grosset}
\address{Department of
Mathematical Sciences,
          Loughborough University,
Loughborough,
          Leicestershire, LE11 3TU, UK}
          \email{M.Grosset@lboro.ac.uk}
          
          %    Information for
%second author
         \author{A.P. Veselov}
\address{Department of
Mathematical Sciences,
          Loughborough University,
Loughborough,
          Leicestershire, LE11 3TU, UK
          and
Landau Institute for Theoretical Physics, Moscow, Russia}

\email{A.P.Veselov@lboro.ac.uk}

\maketitle
\begin{abstract}
A generalisation of the odd Bernoulli polynomials related to the quantum Euler top is introduced and investigated. This is applied to compute the coefficients of the spectral polynomials for the classical Lam\'e operator.
\end{abstract}

%\tableofcontents

\section{Introduction}
The classical  Bernoulli polynomials can be defined through the generating function
$$\frac{ze^{zx}}{e^z-1}=\sum_{k=0}^{\infty}\frac{  B_{k}(x)}{k!}z^k:$$
$$ B_0(x)=1,\qquad   B_1(x)=x-\frac{1}{2},\qquad
B_2(x)=x^2-x+\frac{1}{6},\qquad
B_3(x)=x^3-\frac{3x^2}{2}+\frac{x}{2}, ... $$
(see e.g. \cite{Maths functions, Er}).
They appear naturally in the calculation of sums of powers of the natural numbers
$S_{k-1} (n) = 1^{k-1} + 2^{k-1} +\dots +n^{k-1}$   in a simple way:
$$S_{k-1}(n) = (B_{k}(n+1)-B_{k})/k,$$
where $B_k = B_k(0)$ are the {\it Bernoulli
numbers}. All odd Bernoulli numbers except $B_1= -\frac{1}{2}$ are known to be zero, so the odd Bernoulli polynomials (up to a multiple $k$) can be thought of as an  "analytic continuation" of the sums of powers from natural argument $n$ to real (or complex) $x.$

In this paper we introduce a new class of polynomials, which can be considered as an elliptic generalisation of the {\it odd} Bernoulli polynomials $B_{2k+1}(x)$. They are related to the quantum top and to the classical {\it Lam\'e operator} $$L_s= - \frac{d^2}{dz^2} + s(s+1) \wp(z).$$
where $\wp$ is the Weierstrass elliptic function \cite{Er},
%\[ \wp(z) = \frac{1}{z^2} + \sum_{m,n} ^{ \hspace{0.4 in} {\prime}} \left(\frac{1}{(z- (2n w_1 + 2m w_2))^2} %- \frac{1}{(2n w_1 + 2m w_2) ^2}\right),\] 
satisfying the differential equation
$$(\wp')^2 = 4 \wp^3 - g_2 \wp - g_3.$$
It is well-known after Ince  \cite{Ince} that the Lam\'{e} operator (considered on the real line shifted by the imaginary half-period) for integer $s$ has a remarkable property: its spectrum has exactly $s$ gaps.
The ends of the spectrum  $E_j$ correspond to the doubly periodic solutions of the Lam\'e equation (so-called {\it Lam\'e functions}). We will call the corresponding polynomials $R_{2s+1}(E) = \prod_{j=0}^{2s} (E - E_j(s))$ {\it Lam\'e spectral polynomials.} The computation of the polynomials  $R_{2s+1}(E)$ for given $s=1,2,3,....$ goes back to Hermite and Halphen \cite{Whittaker}. In more recent time this was investigated within the finite-gap theory initiated by S.P. Novikov's work \cite{Nov1} (see \cite{BE, Take1, Take2} for the latest results in this direction). 

Here we consider a related but different problem: we would like to express the coefficient $b_k$ of the spectral polynomial $R_{2s+1}(E) = E^{2s+1} + b_1 E^{2s} + b_2 E^{2s-1}+ \dots + b_{2s+1}$ as a {\it function of $s$} (and thus for {\it all} values of parameter $s$). We will show that in this relation naturally appear some new polynomials, generalising the  odd Bernoulli polynomials. 

The following remarkable relation between the Lam\'e equation and the quantum Euler top, going back to Kramers and Ittmann \cite{Kramers}, will be crucial for us. Consider the quantum mechanical Hamiltonian of the Euler top  (see e.g.  \cite{Landau Q M})
\[ \hat{H} = a_1 { \hat {M}_1}^2 + a_2{\hat {M}_2}^2+a_3 {\hat {M}_3}^2,\]
where  the angular momentum operators $\hat M_j$ satisfy the standard commutation relations $[\hat M_1, \hat M_2] = i \hat M_3, [\hat M_2, \hat M_3] = i \hat M_1, [\hat M_3, \hat M_1] = i \hat M_2$ (we assume $\hbar =1$ for simplicity).

The operator $\hat H$ naturally acts in any representation of the Lie algebra $so(3).$ In particular, it acts in the representation space with spin $s$ of dimension $2s+1$ as a finite-dimensional operator $\hat H_s.$ The claim is that if the parameters $a_i = e_i$ are the roots  $e_1, e_2, e_3$ of the equation
$4 \wp^3 - g_2 \wp - g_3=0,$ then the characteristic polynomial of the operator $\hat{H}_s$ coincides with the spectral Lam\'e polynomial:
\begin{equation}
\label{charH}
 \det (\lambda I - \hat H_s) = R_{2s+1}(\lambda).
\end{equation}  
%In other words, the spectrum of the Euler quantum top in representation with integer spin $s$ %coincides with the spectral ends of the Lam\'e operator.
We discuss this in more detail in the next section.

The Weierstrass condition $e_1 + e_2 + e_3 =0$ is unnatural from this point of view (and moreover contradicts the "physical" condition of positivity of $a_i$), so we consider the case when the parameters $a_i$ are arbitrary. Let us introduce new parameters $g_1, g_2, g_3,$ which are symmetric functions of $a_1, a_2, a_3$ defined by the relation
\begin{equation}
\label{gs}
4(z-a_1)(z-a_2)(z-a_3) = 4 z^3 - g_1 z^2 - g_2 z - g_3.
\end{equation}
We define the {\it  elliptic Bernoulli polynomials} $\mathcal {B}_{2k+1}$
as the coefficients in the expansion of the trace of the resolvent of $\hat H_s$ at infinity:
\begin{equation}
\label{resolv}
tr (\lambda I - \hat H_s)^{-1} = \sum_{k=0}^{\infty} \frac{\mathcal {B}_{2k+1}(s)}{\lambda^{k+1}}
\end{equation}
or, equivalently by the relation 
$$\mathcal {B}_{2k+1}(s) = tr \hat H^k_s.$$ 
$\mathcal {B}_{2k+1}$ is a polynomial in $s$ of degree $2k+1$ with the coefficients, which are polynomials in $g_1, g_2, g_3$ with rational coefficients.  Strictly speaking we should write $\mathcal {B}_{2k+1}(s; g_1, g_2, g_3)$ rather than $\mathcal {B}_{2k+1}(s),$
but we will use both notations depending on the context. When $g_2 = g_3 = 0$  these polynomials reduce up to a factor to the classical odd Bernoulli polynomials: 
$$
\mathcal {B}_{2k+1} (s; g_1, 0, 0)=\frac{g_1^k}{(2k+1)2^{2k-1}}B_{2k+1}(s+1).
$$
%In the Appendix one can find the explicit form of the first 8 elliptic Bernoulli polynomials.
The corresponding elliptic curve $\Gamma$ given by the equation $$y^2 = 4 x^3 - g_1x^2 - g_2 x -g_3$$ degenerates to a rational curve in this case. 
If $g_1=0$ we have the standard Weierstrass form of an elliptic curve. The polynomials $\mathcal {B}_{2k+1} (s; 0, g_2, g_3)$ are called {\it reduced elliptic Bernoulli polynomials} and denoted as $\mathcal {B}^W_{2k+1}(s; g_2, g_3)$ (W is for Weierstrass):  $\mathcal {B}^W_1=2s+1, \quad \mathcal {B}^W_3=0$,
$$\mathcal {B}^W_5= 
\frac{1}{60}g_2s(s+1)(2s-1)(2s+1)(2s+3),
$$
$$\mathcal {B}^W_7=
\frac{1}{280}g_3s(s+1)(2s-3)(2s-1)(2s+1)(2s+3)(2s+5),
$$
$$\mathcal {B}^W_9= 
\frac{1}{1680}g_2^2s(s+1)(2s-1)(2s+1)(2s+3)(4 s^4 + 8s^3 -11 s^2 - 15s +21).
$$

The coefficients of  $\mathcal {B}^W_{2k+1}$ are homogeneous polynomials in $g_2, g_3$ of weight $2k$ if we assume as usual that the weights of $g_2$ and $g_3$ are 4 and 6 respectively (in other words, they are modular forms of weight $2k$, see e.g. \cite{Lang}). Two interesting special cases
$g_2=0$ and  $g_3=0$ are called {\it lemniscatic} and  {\it equianharmonic} respectively and correspond to elliptic curves with additional symmetries.

We will present some effective ways to compute the elliptic Bernoulli polynomials, investigate their properties and then apply them to the calculation of the coefficients of the Lam\'e spectral polynomials. In particular we prove that the coefficient $b_k= b_k(s)$ of the Lam\'e spectral polynomial $R_{2s+1}(E) = \prod_{j=0}^{2s} (E- E_j(s)) = E^{2s+1} + b_1 E^{2s}+ b_2 E^{2s-1}+ ...+ b_{2s+1} $ is a polynomial in $s, g_2, g_3$ with rational coefficients. It can be computed using the reduced elliptic Bernoulli polynomials  by the following recurrence relation with $b_0=1$:
\[ b_k =- \frac{1}{k}\sum_{j=1}^{k} \mathcal {B}^W_{2j+1}(s) b_{k-j}.\]  
The first coefficients are $b_1 =0, \quad b_2= -\frac{g_2}{120}s(s+1)(2s-1)(2s+1)(2s+3),$
\[b_3= - \frac{g_3}{840}s(s+1)(2s-3)(2s-1)(2s+1)(2s+3)(2s+5),\]
\[ b_4 =  \frac{g_2^2}{201600}s(s-1)(s+1)(2s-1)(2s+1)(2s+3)(56s^4 +76 s^3 -94s^2+201s+630)\] 
(see more of them below in section 5). 

Note that once the coefficients $b_k(s)$ are known for $k=0, 1, ...,  2s$ one can find the eigenvalues of the quantum Euler top in the representation with spin $s$ (integer or half-integer) by solving the corresponding algebraic equation $R_{2s+1}(E) = 0.$

We conclude with the discussion of possible relations and further developments.

\section{Lam\'{e} equation and quantum Euler top \label{Quantum top}}

The observation that the Lam\'e equation is closely related to the quantum top was done by Kramers and Ittmann at the early age of quantum mechanics  \cite{Kramers} (see also \cite{Wang}). They showed that the corresponding Schr\"{o}dinger  equation is separable in the elliptic coordinate system and that the resulting differential equations are of Lam\'{e} form. We are going here to re-derive this result and reformulate it in the modern terms.

Consider the Hamiltonian
\[ \hat{H} = a_1 { \hat {M}_1}^2 + a_2{\hat {M}_2}^2+a_3 {\hat {M}_3}^2\]
acting in the space of functions on the unit sphere 
\begin{equation}
q_1^2 + q_2 ^2 + q_3 ^2 =1,
\label{sphere}
\end{equation}
using the standard representation of the angular momenta as the first order differential operators 
\[\hat{M}_1= - \imath (q_2 \partial_{3} - q_3 \partial_{2})\]
\[\hat{M}_2= - \imath (q_3 \partial_{1} - q_1 \partial_{3})\]
\[\hat{M}_3=- \imath ( q_1 \partial_{2} - q_2 \partial_{1}).\]

Let us introduce the {\it elliptic} (or {\it sphero-conical}) coordinates $u_1, u_2$ on this sphere as the roots of the quadratic equation
\begin{equation}
 \frac{q_1 ^2}{a_1-u}+ \frac{q_2 ^2}{a_2-u}+ \frac{q_3 ^2}{a_3-u}=0,
 \label{elliptic cartesian coord2}
\end{equation}
where the parameters $a_1, a_2, a_3$ are the same as in the top's Hamiltonian.
One has then the following expressions for the cartesian coordinates in terms of $u_1, u_2$:
\begin{eqnarray}
q_1^2 &=& \frac{(a_1-u_1)(a_1-u_2)}{ (a_1-a_2)(a_1-a_3)} \nonumber\\
q_2^2 &=& \frac{(a_2-u_1)(a_2-u_2)}{ (a_2-a_1)(a_2-a_3)} \label{elliptic cartesian coord1}\\
q_3^2 &=& \frac{(a_3-u_1)(a_3-u_2)}{ (a_3-a_1)(a_3-a_2)}. \nonumber
\end{eqnarray}

The system has an obvious quantum integral (Casimir) $\hat{M}^2= \sum \hat{M_i}^2,$ which is the square of the total angular momentum operator:
\[[\hat{M}^2, \hat{M_i}]=0.\] 
%This follows immediately from the commutation relations
 One can check that that in the elliptic coordinate system the operators $\hat H$ and $\hat{M}^2$ have the form
\begin{equation}
\hat{M}^2 = - \frac{4}{u_1-u_2}[\sqrt{-P(u_1)}\frac{\partial}{\partial u_1}(\sqrt{-P(u_1)} \frac{\partial}{\partial u_1})+\sqrt{P(u_2)}\frac{\partial}{\partial u_2}(\sqrt{P(u_2)} \frac{\partial}{\partial u_2}) ]
\label{Casimir}
\end{equation}
\begin{equation}
\hat{H} = - \frac{4}{u_1-u_2}[u_2\sqrt{-P(u_1)}\frac{\partial}{\partial u_1}(\sqrt{-P(u_1)} \frac{\partial}{\partial u_1})+ u_1\sqrt{P(u_2)}\frac{\partial}{\partial u_2}(\sqrt{P(u_2)} \frac{\partial}{\partial u_2}) ]
\label{Hamiltonian}
\end{equation}
where $P(u)=(u-a_1)(u-a_2)(u-a_3).$
Note that the operator $\hat{M}^2$ corresponds to the standard Laplacian $-\Delta$ on the unit sphere.

Since $\hat{M}^2$ and $\hat{H}$ commute, one can look for joint eigenfunctions. The spectral problem $\hat{M}^2 \psi = \mu \psi$ is well-known in the theory of spherical harmonics (see e.g. \cite{mueller}).
It is known that the spectrum has the form $\mu = s(s+1)$ for non-negative integer values of $s$. The dimension of the corresponding eigenspace $V_s$ is $ 2s+1$ and $V_s$ is an irreducible representation of dimension $2s+1$ of the rotation group $SO_3$ called {\it representation with spin $s$.}

It turns out that the joint eigenvalue problem 
\[ \hat{M}^2 \phi = s(s+1) \phi \]
\[ \hat{H} \phi = E \phi\]
is separable in the elliptic coordinates $u_1, u_2$ (see \cite{Kramers, Wang}).
Namely, if we substitute $\phi (u_1,u_2)=\phi_1(u_1)\phi_2(u_2)$ into this system we find that each of the functions $\phi_1(u_1),\phi_2(u_2)$ satisfies the same differential equation:
\[(4[P(u)]^\frac{1}{2} \frac{d}{du}([P(u)]^\frac{1}{2} \frac{d}{du}) -s(s+1)u  +E)\psi =0,\]
which can be rewritten as
\begin{equation}
\frac{d^2}{{du}^2}\psi + \frac{1}{2}[\frac{1}{u-a_1} +\frac{1}{u-a_2}+\frac{1}{u-a_3}]\frac{d}{du}\psi=\frac{1}{4}\frac{s(s+1)u -E}{(u-a_1)(u-a_2)(u-a_3)}\psi.
\label {Lame algebraic form}
\end{equation}

A remarkable fact is that this is an algebraic form of the following slightly generalised version of the  Lam\'{e} differential equation
\begin{equation}
\label{Lame}
 - \frac{d^2}{dz^2}\psi + s(s+1) \wp_*(z)\psi = E \psi
\end{equation}
where $\wp_*(z)$ is a solution of the differential equation
\begin{equation}
\label{wp*}
 (\wp'_*)^2 = 4( \wp_*-a_1) ( \wp_*-a_2)( \wp_*-a_3).
 \end{equation}
Indeed, after the change of variables $u = \wp_*(z)$ the equation (\ref{Lame}) coincides with (\ref{Lame algebraic form}) (see \cite{Whittaker}). When the sum $a_1+a_2+a_3=0$ the equation (\ref{wp*}) determines the Weierstrass elliptic function $\wp(z)$, otherwise it differs from it by adding a constant.

It is well-known (see e.g. \cite{Er}) that for $\phi$ to be a regular solution on the sphere the corresponding $\psi$ must be doubly-periodic, which implies that $s$ is integer and $E$ must have one of the $2s+1$ characteristic values $E_m(s)$. For each $E_m(s)$ there exists exactly one (up to a factor)  doubly-periodic solution to the Lam\'{e} equation $\mathcal {E}_s^m(u),$ which is called the {\it Lam\'{e} function}. Therefore the basis of the eigenfunctions of the operator $\hat H$ in the invariant subspace $V_s$ consists of  $2s+1$ solutions $\phi (u_1,u_2)$ of the form $\mathcal {E}_s^m(u_1) \mathcal {E}_s^m(u_2).$ They are called sometime {\it ellipsoidal harmonics} (see \cite{Whittaker}).

Thus, we come to the following result (cf. \cite{Kramers, Wang}):
 
 \begin{theorem} The characteristic polynomial of the quantum top Hamiltonian  $\hat{H}_s$ in the representation space with integer spin $s$ coincides with the spectral polynomial $R_{2s+1}(\lambda) = \prod_{j=0}^{2s} (\lambda - E_j(s))$ of the generalised Lam\'{e} operator (\ref{Lame}).
% $$R_{2s+1}(\lambda) = \det (\lambda I - \hat H_s).$$
 \end{theorem}

{\it Remark 1.} Turbiner \cite{Turb} has discovered a similar but different relation of the Lam\'e equation with certain quadratic elements of the universal enveloping $sl(2).$ The Lam\'e spectral polynomials are known to be factorisable and Turbiner's result gives an interesting interpretation for the factors in these terms.

{\it Remark 2.} A simple relation between the quantum Euler top and the Lam\'e equation mentioned above is a bit misleading. Indeed there are several spectral problems related to the Lam\'e equation. We have considered only smooth real periodic version related to real $x$ shifted by the imaginary half-period. If we would consider $x$ just real, we would have a singular version (since $\wp$ has poles on the real line), whose spectrum has nothing to do with the quantum top. In its turn, the quantum Euler top in the representation with half-integer spin $s$ has eigenvalues which are just some special double eigenvalues of the periodic Lam\'e operator, which in this case has infinitely many gaps.

\section{Elliptic Bernoulli polynomials}

We define now the {\it elliptic Bernoulli polynomials} $\mathcal {B}_{2k+1}(s)$ as the traces of the powers of $\hat H_s,$ where $\hat H_s$ is as before the quantum top operator $\hat{H}$ in the representation with spin $s$:
\begin{equation}
\label{ellB}
\mathcal {B}_{2k+1}(s; g_1,g_2, g_3) = tr \hat H_s^k, \quad k=0,1,2,\dots
\end{equation}
Here the parameters $g_1= 4(a_1 + a_2 +a_3), g_2 = -4(a_1a_2 + a_2 a_3 + a_1 a_3),
g_3 = 4a_1 a_2 a_3$ are defined by the relation (\ref{gs}).

\begin{theorem} The trace $tr \hat H_s^k$ is a polynomial in $s$ of degree $2k+1$ anti-symmetric with respect to $s=-\frac{1}{2}$, whose coefficients are polynomials in $g_1, g_2, g_3$ with rational coefficients.  When $g_2 = g_3 = 0$  it reduces (up to a factor and shift) to the corresponding classical odd Bernoulli polynomial:
\begin{equation}
\mathcal {B}_{2k+1} (s; g_1, 0, 0)=\frac{g_1^k}{(2k+1)2^{2k-1}}B_{2k+1}(s+1).
\label{general trace hyp}
\end{equation} \end{theorem}

The first part essentially follows from the Harish-Chandra general results \cite{HCh} (see also \cite{D}, page 268), but we give here a direct proof. 

Consider the standard basis  in $V_s$ consisting of the eigenvectors $|j>$ of $\hat {M}_3$: $\hat {M}_3 |j>  = j |j>, \, j= -s, -s+1, \dots s-1, s.$ In this basis, the Hamiltonian $\hat H$ is a tri-diagonal  symmetric matrix $H=H_s$ with the following elements (see e.g. Landau-Lifshitz \cite{Landau Q M},  page 417):
\begin{equation}
<j|H|j> = \frac{1}{2} (a_1+a_2) [ s(s+1) - j^2] + a_3 j^2
\label{Hamiltonian - diagonal element}
\end{equation}
\[<j|H|j+2> = < j+2 | H |j> = \frac{1}{4} (a_1-a_2) \sqrt{ (s-j)(s-j-1)(s+j+1)(s+j+2)}.\] 
Note that both expressions are symmetric with respect to $s=-\frac{1}{2};$ they are also homogeneous polynomials of degree 1 in $a_1,a_2, a_3$. Now, consider any diagonal element of $H^k;$ it has the form:
\[< j|H^k|j> = \sum_{i_1,i_2,...,i_{k-1}} <j|H|i_1> <i_1|H|i_2>...... <i_{k-1}|H|j>\]  where the distance between 2 consecutive indices $i_l, i_{l+1}$ is either 0 or $\pm 2$. Since the starting point and the ending point coincide, if the matrix element $<i_l|H|i_l+2>$ appears along the path so does the element $ < i_l+2 | H |i_l>.$ This proves that  the diagonal matrix elements of $H^k$ are polynomials of degree $2k$ in both $s$ and $j$. From (\ref{Hamiltonian - diagonal element}) they are symmetric with respect to $s=-\frac{1}{2}$ and homogeneous symmetric polynomials of degree $k$ in $a_1, a_2, a_3.$  
Now summing over $j= -s, -s+1, \dots s-1, s$ and taking into account that the sums of the odd powers of $j$ are zero while the sums of even powers $2l$ are the odd Bernoulli polynomials $B_{2l+1}(s+1)$ (multiplied by $\frac{2}{2l+1}$) we have the first statement of the theorem. The anti-symmetry of $\mathcal {B}_{2k+1}(s)$ with respect to $s=-\frac{1}{2}$  follows from the well-known property of the Bernoulli polynomials: $B_{m}(1-s) = (-1)^m  B_{m}(s).$

In the case when $a_1=a_2=0,$ we have $g_2=g_3=0, ~ g_1 = 4 a_3$ and $\hat H = a_3 \hat M_3^2.$
The spectrum of $H_s$ is then very simple: $\lambda_j = a_3 j^2$ for $ j= -s, -s+1, \dots, s-1, s.$
Since the sum $\sum_{j=1}^{s}  j^{2k} = \frac{1}{2k+1} B_{2k+1}(s+1),$ we thus obtain (\ref{general trace hyp}). This completes the proof of Theorem 1.

Note that from the point of view of the elliptic curve $\Gamma$ given by the equation $$y^2 = 4 x^3 - g_1x^2 - g_2 x -g_3,$$ the last case corresponds to the limit when one of the periods goes to infinity
("trigonometric limit"). There are two more interesting special cases:
 {\it lemniscatic} when $g_1 =g_3 =0$ and {\it equianharmonic} when $g_1=g_2=0,$ 
 corresponding to the elliptic curves with additional symmetries.

 It is natural also to consider the Weierstrass reduction $g_1=0;$ we will call the corresponding polynomials 
 $ \mathcal {B}^W_{2k+1} (s; g_2, g_3) = \mathcal {B}_{2k+1} (s; 0, g_2,g_3)$ the {\it reduced elliptic Bernoulli polynomials.}

\begin{theorem} The elliptic Bernoulli polynomial $\mathcal {B}_{2k+1}$  has the following properties:
\begin{enumerate}
\item as a polynomial in $g_1, g_2, g_3$ $\mathcal {B}_{2k+1}$ is homogeneous of weight $2k,$ where the weights of $g_1, g_2$ and $g_3$ are assumed to be 2, 4 and 6 respectively, 
\item $\mathcal {B}_{2k+1}$ for $k \geq 1$ is divisible by $s(s+1)(2s+1),$
\item in the reduced case $\mathcal {B}^W_{2k+1}$ is divisible by $s(s+1)(2s-1)(2s+1)(2s+3)$ for all $k$ and  by $s(s+1)(2s-1)(2s+1)(2s+3)(2s-3)(2s+5)$ for odd $k,$ 
\item in the lemniscatic case  $\mathcal {B}_{2k+1}(s; 0, g_2, 0)=0$ for odd integer $k,$
\item in the equianharmonic case $\mathcal {B}_{2k+1}(s; 0, 0, g_3)=0$ if $k$ is not divisible by 3,
\item in the isotropic case $a_1=a_2=a_3=a$ i.e  $g_1=12a, g_2=-12a^2, g_3=4a^3,$
$\mathcal {B}_{2k+1}(s) = a^k (2s+1)s^k(s+1)^k.$
\end{enumerate}
\end{theorem}

The proof of the first two claims follows from the definition and the anti-symmetry property. To prove the third one consider the representation with spin $ s=\frac{1}{2}.$ It is easy to check that  $\hat{H}$ acts as the 2 by 2 scalar matrix $\frac{1}{4}(a_1+a_2+a_3) Id,$ which is zero in the reduced case. Therefore $\mathcal {B}_{2k+1}^W(\frac{1}{2})=0$ for all $k$. By anti-symmetry with respect to $-\frac{1}{2}$ we also have $\mathcal {B}_{2k+1}^W(-\frac{3}{2})=0.$
For  half integer $s$, we know from Kramers' theorem (see \cite{Landau Q M}, paragraph 60) that the eigenvalues are no longer distinct but are double roots. For the particular case $s= 3/2,$ these eigenvalues take the values $\pm \sqrt{[3(a_1^2 + a_2^2 + a_3^2)/2]}$ (see \cite{Landau Q M}, page 419) therefore  for odd $k,$ $\mathcal {B}_{2k+1}^W(3/2)=0$  and again by anti-symmetry $\mathcal {B}_{2k+1}^W(-5/2)=0.$
The lemniscatic and equianharmonic cases  follow from the first claim. In the isotropic case
$\hat H_s = a s(s+1) Id,$ which implies the last statement.

In the general case the elliptic Bernoulli polynomials are not zero and their highest coefficients are described by the following

\begin{theorem} The leading term of the elliptic Bernoulli polynomial $\mathcal {B}_{2k+1}(s) = A_0 s^{2k+1} + A_1 s^{2k} + \dots +A_{2s}$ can be written  
\begin{equation}
\label{formula}
 A_0 s^{2k+1} = 2\int_{0}^{s} Res \hspace{0.1in} \xi^{-1}[ \gamma (s^2 -j^2) \xi + ( \alpha s^2 + \beta j^2) + \gamma ( s^2 - j^2)\xi^{-1}]^k dj,
\end{equation}
 where $\alpha =  \frac{1}{2} (a_1+a_2)$, $\beta = \frac{1}{2} (2a_3 - a_1 - a_2), \, \gamma = \frac{1}{4} (a_1-a_2).$
\label{coeff of highest term}\end{theorem}

Indeed, for large $s$ and $j$ the leading behaviour of the matrix elements of $\hat H$ is
\[ <j|\hat H|j> = \frac{1}{2} (a_1+a_2) [ s^2 - j^2] + a_3 j^2= \alpha s^2 + \beta j^2,\]
\[<j|\hat H|j+2> = < j+2 | \hat H |j> = \frac{1}{4} (a_1-a_2)(s^2 -j^2) = \gamma (s^2 - j^2).\]
Therefore the leading term of the diagonal element $ <j|\hat H^k|j> $ coincide with the constant term of the Laurent polynomial $[\gamma (s^2 -j^2) \xi + ( \alpha s^2 + \beta j^2) + \gamma ( s^2 - j^2)\xi^{-1}]^k$
in auxiliary variable $\xi.$ Replacing the summation over $j$ by the integration, which is fine in the leading order, we come to our formula.

Note that from this formula the fact that the final result is a symmetric function of $a_1, a_2, a_3$
(and thus is a polynomial in $g_1, g_2, g_3$ ) is not obvious at all.

{\it Remark.} From the quasi-classical arguments we can write the highest coefficient $A_0$ 
as the following integral over the unit sphere
\begin{equation}
\label{qformula}
 A_0  = \frac{1}{2\pi} \int_{|M|^2=1} H^k d\Omega  = \frac{1}{2\pi} \int_{|M|^2=1} (a_1 M_1^2 + a_2 M_2^2 + a_3 M_3^2)^k d\Omega,
 \end{equation}
where $d\Omega$ is the area element on the unit sphere. Thus our formula (\ref{formula}) gives an
expression for this integral. It would be interesting to compare it with the calculation of this integral using elliptic coordinates.

\section{Effective way to compute the elliptic Bernoulli polynomials}

Although the definition of the elliptic Bernoulli polynomials themselves gives a way to compute them 
as traces of powers of the given matrices $H_s,$ it does not seem to be as effective as the following procedure based on the fact that the matrix $H_s$ is tri-diagonal.

Indeed, in the standard basis $|j>$ of the space $V_s$ the eigenvalue problem $\hat{H}\psi= \lambda \psi$  leads to the following difference equation: 
\begin{equation}
c_{n-2} \psi_{n-2} + v_n \psi_n + c_n \psi_{n+2}= \lambda \psi_n,
\label{QT difference eq1}
\end{equation}
where $$c_n= \frac{a_1-a_2}{4} \sqrt{(s-n)(s-n-1)(s+n+1)(s+n+2)},$$
$$ v_n= \frac{1}{2} (a_1+a_2)^2 [s(s+1) - n^2] + a_3 n^2.$$ 

For such an equation one can use the standard procedure (see e.g. \cite{FT}) from the theory of solitons to find the local spectral densities, which are difference analogues of the famous KdV densities \cite{Nov}. In our case it works as follows.

Let $\chi_n = \frac{c_n \psi_{n+2}}{\psi_n},$ then the equation (\ref{QT difference eq1}) becomes
\begin{equation}
c_{n-2}^2 + ( v_n - \lambda)\chi_{n-2} + \chi_n \chi_{n-2}=0
\label{QT difference eq2}
\end{equation}
We look for a solution in the form $\chi_n= \lambda - \sum_{k=0}^{\infty} \chi_{n,k} \lambda^{-k}.$ Substitution of this expression into the equation (\ref{QT difference eq2}) gives 
$\chi_{n,0} = v_n, \chi_{n,1} = c_{n-2}^2, \chi_{n,2} = c_{n-2}^2 v_{n-2},$
and for general $k \geq 1$ the recurrence relation:
\begin{equation}
\label{recurr}
\chi_{n,k+1} =  \sum_{i=1}^{k} \chi_{n,i} \chi_{n-2,k-i}.
\end{equation}

Let $X=  \sum_{k=0}^{\infty} \chi_{n,k}\lambda^{-(k+1)}$ so that  $ \chi_n= \lambda (1 - X)$ and $\log  \chi_n= \log \lambda  - \sum_{i=1}^{\infty} \frac{X^i}{i}.$ Thus we have
\begin{equation}
\log \chi_n- \log \lambda = -\sum_{i=1}^{\infty} \frac{\mathcal{I}_{n,i}}{\lambda^i},
\label{QT3}
\end{equation}
where $\mathcal{I}_{n,1}= v_n,$ $\mathcal{I}_{n,2}= c_{n-2}^2 + \frac{v_n^2}{2},$ $\mathcal{I}_{n,3}= c_{n-2}^2 v_{n-2}+  v_n c_{n-2}^2 + \frac{v_n^3}{3},.......$

On the other hand one can check that $\prod_n\frac{ \chi_n }{\lambda} = \prod_m(1-\frac{E_m(s)}{\lambda})$ where the $E_m(s)$ are the eigenvalues of $\hat H_s.$ Thus 
\[\sum_n (\log \chi_n - \log \lambda) = -\sum_n \sum_{i=1}^{\infty} \frac{\lambda_n^i}{i\lambda^i}
=\sum_{i=1}^{\infty} \frac{Tr \hat H_s^i}{i\lambda^i.}\]
Comparing this with the equations (\ref{QT3}), we obtain
\[Tr \hat H_s^k= k\sum_n \mathcal{I}_{n,k}= k \sum_{n=-s}^{s} \mathcal{I}_{n,k} .\]

\begin{theorem} 
The elliptic Bernoulli polynomials $\mathcal {B}_{2k+1}$ can be computed as
\begin{equation}
\mathcal {B}_{2k+1} = k \sum_{n=-s}^{s} \mathcal{I}_{n,k},
\label{effect}
\end{equation}
where $\mathcal{I}_{n,k}$ are the local densities determined by the relations (\ref{recurr}, \ref{QT3}).
\end{theorem}

This gives a very effective way to compute the elliptic Bernoulli polynomials since the local densities are polynomials in $c_n^2$ and $v_n$ (and hence in $n$) and thus the summation over $n$ can be done with the use of the standard Bernoulli polynomials. 
We had applied this procedure to find the first 10 elliptic 
Bernoulli polynomials using Mathematica (see 8 of them in the Appendix).

\section{Application: coefficients of the Lam\'e spectral polynomials} 
 
We will consider again the generalised version of the Lam\'e operator (\ref{Lame}).
The coefficients $b_k=b_k(s)$ of the corresponding spectral polynomial $$R_{2s+1}(E) = \prod_{i=0}^{2s} (E- E_i)= E^{2s+1} + b_1 E^{2s}+ b_2 E^{2s-1}+ ..+  b_k E^{2s - k +1} +...+ b_{2s+1}$$ 
up to a sign are the elementary symmetric functions of the eigenvalues:
$b_k = (-1)^k e_k,$ where $e_1 = \sum E_i, \, e_2 = \sum_{i<j} E_i E_j, \, e_3 = \sum_{i<j<k} E_i E_j E_k... $
The elementary symmetric functions are related to power sums $\mathcal {B}_{2k+1}(s)= \sum E_i ^k$  by the following well-known relations:
\[k e_k = \sum_{j=1}^{k} (-1)^{j-1} \mathcal {B}_{2j+1} e_{k-j}\] with $e_0 =b_0=1$ 
(see e.g. \cite{Symmetric functions}). 
This implies the following 

\begin{theorem} The coefficients $b_k$ of the Lam\'e spectral polynomial $R_{2s+1}(E)$
%$$R_{2s+1}(E) = \prod_{i=0}^{2s} (E- E_i(s)) = E^{2s+1} + b_1 E^{2s}+ b_2 E^{2s-1}+ ...+ b_{2s+1} $$ 
are related to the elliptic Bernoulli polynomials $\mathcal {B}_{2j+1}(s)$ by the recurrent relations
\[ b_k =- \frac{1}{k}\sum_{j=1}^{k} \mathcal {B}_{2j+1}(s)  b_{k-j}.\]  
The coefficient  $b_k$ is a polynomial in $s, g_1,g_2, g_3$ with rational coefficients. As a polynomial in $s$ it has degree $3k$ and is divisible by $(s+1)s(s-1)...(s-[\frac{k-2}{2}]).$
\end{theorem}

One can apply this result also to the case of half-integer spin $s$: in that case all the roots of the polynomial $R_{2s+1}(E)$ are double and correspond to the doubly-periodic solutions of the Lam\'e equation.

In the reduced case ($g_1=0$) the degree of $b_k$ drops to $[\frac{5k}{2}]$ (for $k>1$).
Using the explicit form of the elliptic Bernoulli polynomials given in the Appendix one can find the first seven coefficients $b_k$, which in reduced case are: $b_1 =0,$
\[b_2= -\frac{g_2}{120}s(s+1)(2s-1)(2s+1)(2s+3)\]
\[b_3= - \frac{g_3}{840}s(s+1)(2s-3)(2s-1)(2s+1)(2s+3)(2s+5)\]
\[ b_4 =  \frac{g_2^2}{201600}s(s-1)(s+1)(2s-1)(2s+1)(2s+3)(56s^4 +76 s^3 -94s^2+201s+630)\] 
\small
\[ b_5 =+\frac{g_2g_3}{1108800}s(s-1)(s+1)(2s-3)(2s-1)(2s+1)(2s+3)(2s+5)(88 s^4 +68 s^3-302 s^2+663 s + 1890) \]
\normalsize
\[ b_6 = \frac{g_3^2}{201801600} (s-2)(s-1)s(s+1)(2s-3)(2s-1)(2s+1)(2s+3)(2s+5) \times \] \[(4576 s^5+12944 s^4-20720 s^3+48312 s^2+597150 s+779625) -\]  \[\frac{g_2^3}{10378368000}(s-2)(s-1)s(s+1)(2s-5)(2s-3)(2s-1)(2s+1)(2s+3)\times\] \[(16016 s^6+89232 s^5+197160 s^4+544280 s^3+2033829 s^2+3858813 s+2619540)\]
\[ b_7 =  -\frac{g_2 ^2 g_3}{24216192000}(s-3)(s-2)(s-1)s(s+1)(2s-5)(2s-3)(2s-1)(2s+1)(2s+3)(2s+5) \times\] \[ (32032 s^6 +189072 s^5+463440 s^4+1682920 s^3+7301418 s^2+15249213 s+11351340)\]

\section{Concluding remarks.}
 
We have shown that for any given $k$ the coefficient $b_k (s)$ of the spectral Lam\'e polynomial $R_{2s+1}$ can be computed effectively for all values of parameter $s.$ In particular, for fixed $s$ it gives an alternative way to compute the whole polynomial. It would be interesting to compare this approach with the classical one going back to Halphen and Hermite \cite{Whittaker} further developed recently by Belokolos and Enolski \cite{BE} and Takemura \cite{Take1, Take2} following the work of Krichever \cite{Krich}.

However we believe that  the elliptic Bernoulli polynomials are of interest by themselves. In particular one can expect interesting relations with the arithmetic of the corresponding elliptic curves and the representation theory. In this relation we would like to mention the elliptic generalisation of the Bernoulli numbers - the so-called {\it Bernoulli-Hurwitz numbers} $BH_{2k}$,
whose arithmetic was investigated in \cite{Katz, Onishi}. 

Another interesting possible relation is with the zeta-function $\zeta_H (z) = tr \hat H^{-z}$ of the quantum top and its special values. A lemniscatic case $a_3 = \frac{a_1+a_2}{2}$ could be particularly interesting from the arithmetic point of view.

Recall that the parameter $s$ was originally integer or half-integer (spin).
A natural question is the role of these values in the theory of elliptic Bernoulli polynomials. We conjecture that like in the case of the usual Bernoulli polynomials (see e.g. \cite{VW}) these values are the asymptotic positions of the real roots of the polynomials $\mathcal {B}_{2k+1}$ for large $k$. More precisely, we conjecture that for real $s$ in the bounded interval the ratio
$$ \frac{\mathcal {B}_{2k+1}(s)}{\mathcal {B}'_{2k+1}(0)} \rightarrow \frac{\sin 2\pi s}{2 \pi}$$
as $k$ tends to infinity. Actually, we believe that this is true for each component of $\mathcal {B}_{2k+1}$, which is a coefficient at monomial $g_1^p g_2^q g_3^r.$

It is interesting to look at the graphs. In the Fig 1 we show the graphs of the coefficients
of the polynomial $\mathcal {B}_{15}(s) $ at: (a) $g_1^7,$  (b) $g_1^3 g_2^2,$ (c) $g_1^2 g_2 g_3,$  and (d) $g_2^2g_3.$ We normalise each polynomial by dividing it by its first derivative at zero and then multiplying it by $2\pi.$ The sinusoidal behavior for small $s$ looks quite plausible.

\begin{figure}[h]
\centerline{ \includegraphics[width=6cm]{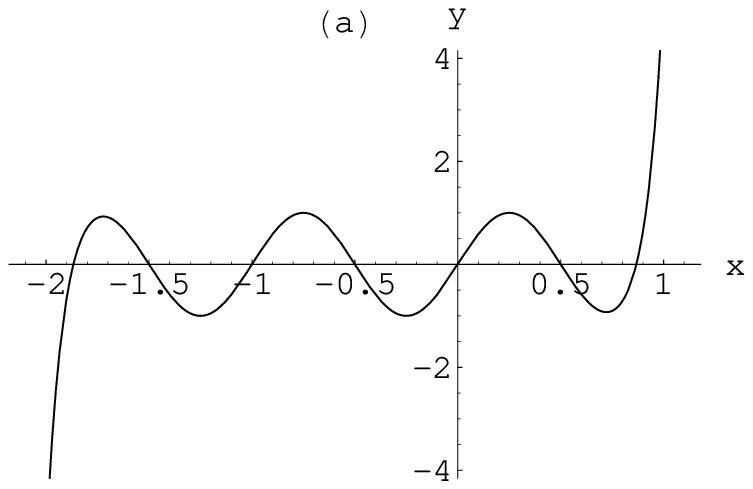} \hspace{10pt}
\includegraphics[width=6cm]{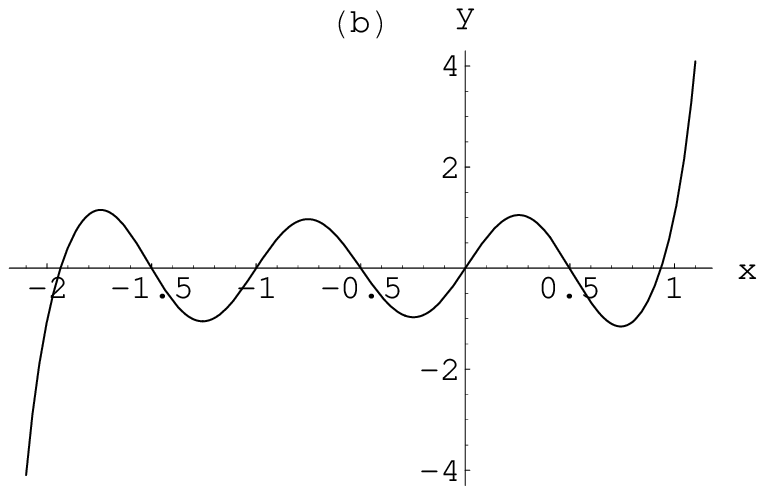} }
\centerline{ \includegraphics[width=6cm] {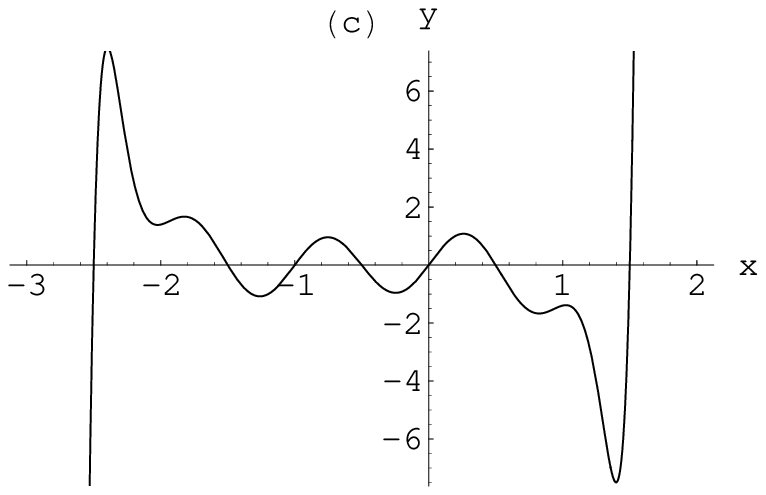}\hspace{10pt}
\includegraphics[width=6cm] {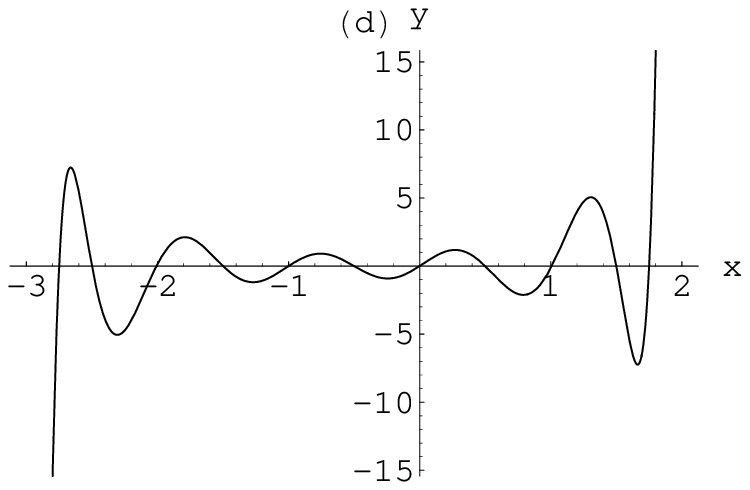}}
\caption{}
\end{figure}

We would like to mention that the even Bernoulli polynomials (or more precisely closely related Faulhaber polynomials) also have elliptic versions related to the Lam\'e operator. They were introduced in our recent paper \cite{GV2} motivated by \cite{FV} as certain complete elliptic integrals of second kind
and have quite different properties. The fact that the theory of the Lam\'e equation leads to two different classes of polynomials, both related to Bernoulli polynomials (one to odd, another to even) seems to be
remarkable. To make the picture even more intriguing we would like to mention that the integrals in the definition of the elliptic Faulhaber polynomials are coming from the formal expansion of the trace of the resolvent of the Lam\'e operator (cf. our formula (\ref{resolv})).

Another interesting problem is to investigate the analogues of elliptic Bernoulli polynomials related to Sklyanin algebra. It is known after Krichever and Zabrodin \cite{KZ} that Sklyanin's representation \cite{Skl} gives certain difference analogue of the Lam\'e equation, so one can consider the traces of powers
of the corresponding generator $S_0$ as functions of the corresponding spin. We would like to mention in this relation a very interesting
paper \cite{Smirn}, where the traces on the Sklyanin algebra are discussed. In particular, the formulas (2.20), (2.21) from this paper give an explicit expression of the traces of $S_0$ and $S_0^2$
in terms of elliptic functions, which show that they are not polynomials anymore.

Finally, one can consider our results from the general point of view of the quantisation of integrable systems.
Usually one can find the spectrum in a closed form only if the classical system is integrable in elementary functions. The Euler top is probably the most natural classical problem integrable in elliptic functions. The question about the nature of its integrability in the quantum case seems to be not as easy as it may look. We hope that our paper adds something in this direction as well.

 \section{Acknowledgements.}

We are grateful to H.R. Dullin, V.Z. Enolski, V.A. Fateev, J. Gibbons, A.R. Its, J. Samson, A.N. Sergeev, E.K. Sklyanin, F. A. Smirnov and A.V. Turbiner for useful and stimulating discussions.

One of us (APV) would like to acknowledge the support of European research
programmes ENIGMA (contract MRTN-CT-2004-5652) and MISGAM.

%\newpage

 \section{Appendix  - The first 8 elliptic Bernoulli polynomials \label{Appendix generalised elliptic Bernoulli}} 
 
\bigskip 
\noindent 
$\mathcal {B}_1  = 
 \begin{array}{llll}
 &2s + 1
 \end{array}$
 \newline \newline
$\mathcal {B}_3  =
 \begin{array}{llll}
 &\frac{1}{12}&{g_1}& s (s+1)(2s+1)
 \end{array}$
 \newline \newline
  $\mathcal {B}_5  = 
  \begin{array}{llll}
  &\frac{1}{240}&{g_1}^2 &s (s+1)(2s+1)(3 s^2 +3s -1)\\
  +&\frac{1}{60}&{g_2}&s(s+1)(2s-1)(2s+1)(2s+3)
  \end{array}$
 \newline \newline \newline \newline
   $\mathcal {B}_7 =
\begin{array}{llll}
&\frac{1}{1344}& {g_1}^3 &s(s+1)(2s+1)(3 s^4 + 6 s^3 -3s +1) \\
+& \frac{1}{1120}&{g_1} {g_2}& s(s+1)(2s-1)(2s+1)(2s+3)(6 s^2 +6s -5)\\
+& \frac{1}{280}&{g_3}&s(s+1)(2s-3)(2s-1)(2s+1)(2s+3)(2s+5)\end{array}
$
 \newline \newline \newline \newline
$\mathcal {B}_9=
\begin{array}{llll}
&\frac{1}{11520}& {g_1}^4& s(s+1)(1+2s)(5 s^6 +15 s^5 +5 s^4 -15 s^3 -s^2 +9s -3) \\ +&\frac{1}{3360}&{g_1}^2 {g_2} & s(s+1)(2s-1)(2s+1)(2s+3)(5 s^4 +10 s^3 -5 s^2 -10 s +7)\\
+&\frac{1}{840}&{g_1} g_3&  s^2(s+1)^2(2s-3)(2s-1)(2s+1)(2s+3)(2s+5)\\ 
+&\frac{1}{1680}&{g_2}^2&s(s+1)(2s-1)(2s+1)(2s+3)(4 s^4 + 8s^3 -11 s^2 - 15s +21)
  \end{array}$
  \newline \newline\newline \newline
  $\mathcal {B}_{11} = \small
\begin{array}{llll}
&\frac{1}{33792}& {g_1}^5&s(s+1)(2s+1)(s^2 + s -1)( 3 s^6 +9 s^5 +2s^4 -11 s^3 +3 s^2 +10 s -5)  \\ +&\frac{1}{50688}&{g_1}^3 g_2 & s(s+1)(2s-1)(2s+1)(2s+3)(20s^6 +60s^5-10 s^4 -120 s^3 +44 s^2 +114 s -75)\\
+&\frac{1}{29568}&{g_1}^2 {g_3}&  s(s+1)(2s-3)(2s-1)(2s+1)(2s+3)(2s+5)( 10 s^4 +20 s^3 -4 s^2 -14 s +21)\\ +&\frac{1}{29568}&{g_1}{g_2}^2&  s(s+1)(2s-1)(2s+1)(2s+3)( 40 s^6 +120 s^5-86 s^4 -372 s^3+242 s^2 +448 s -315) \\
+&\frac{1}{7392}&{g_2}{ g_3}& s(s+1)(2s -3)(2s-1)(2s+1)(2s+3)(2s+5)(8 s^4 + 16s^3 -34 s^2 - 42s +63)
  \end{array}$
  \newline \newline \newline 
  $\mathcal {B}_{13} = \small
 \begin{array}{llll}
&\frac{1}{5591040}& {g_1}^6&s(s+1)(2s+1)( 105 s^{10}+525 s^9+525 s^8-1050s^7-1190 s^6 \\ &&& +2310 s^5 +1420s^4 -3285 s^3 -287 s^2 +2073 s -691)  \\ \\ +&\frac{1}{5125120}&{g_1}^4 {g_2} & s(s+1)(2s-1)(2s+1)(2s+3)\\ &&&(525 s^8+2100s^7+350 s^6 -6300s^5-70 s^4 +12810 s^3 -4105 s^2 -11910 s +7601)\\ \\
+&\frac{1}{2842840}&{g_1}^3 {g_3}&  s(s+1)(2s-3)(2s-1)(2s+1)(2s+3)(2s+5)\\ &&&( 350 s^6+1050 s^5-100 s^4 -1950 s^3 +1433 s^2 +2583 s -1650)\\  \\+&\frac{1}{2562560}&{g_1}^2 {g_2}^2&  s(s+1)(2s-1)(2s+1)(2s+3)(1400 s^8 +5600 s^7-1450s^6-23950 s^5\\ &&&+5438 s^4+57326 s^3-24627 s^2-58215 s +41481) \\ \\
+&\frac{1}{320320}&{g_1}{g_2}{ g_3}& s(s+1)(2s -3)(2s-1)(2s+1)(2s+3)(2s+5)\\ &&&(200s6+600s5-670 s^4 -2340s^3 +1922 s^2 +3192s -2475)\\ \\
+&\frac{1}{960960}&{g_2}^3&s(s+1)(2s-1)(2s+1)(2s+3) (400s^8+1600s^7-1640s^6-10520 s^5 \\ &&&+8193 s^4+35786s^3-28282s^2-48195s+43659)\\ \\ +&\frac {1}{160160}&{g_3} ^2&s(s+1)(2s -3)(2s-1)(2s+1)(2s+3)(2s+5)\\ &&&(80s^6 +240 s^5-840 s^4-2080 s^3+4401 s^2+5481 s-7425)
  \end{array}$
  \normalsize
\newline \newline \newline \newline 
  $\mathcal {B}_{15} = \small
\begin{array}{llll}
&\frac{1}{737280}& {g_1}^7&s(s+1)(2s+1)( 3s^12+18s^11+24 s^{10}-45 s^9-81 s^8+144s^7+182 s^6 \\ &&& -345 s^5 -217s^4 +498 s^3+44 s^2 -315 s+105)  \\ \\+&\frac{1}{1597440}&{g_1}^5 \tilde{g_2} & s(s+1)(2s-1)(2s+1)(2s+3)(42 s^{10}+210 s^9+105 s^8-840s^7\\ &&&-364 s^6 +2730s^5 +205 s^4 -5540 s^3 +1650 s^2+5078 s -3185)\\ \\
+&\frac{1}{13178880}&{g_1}^4 {g_3}&  s(s+1)(2s-3)(2s-1)(2s+1)(2s+3)(2s+5)( 315s^8+1260s^7 \\ &&&+140 s^6-3990 s^5+1265 s^4 +10650 s^3 -5152 s^2 -11352 s +9009)\\ \\+&\frac{1}{13178880}&{g_1}^3 {g_2}^2&  s(s+1)(2s-1)(2s+1)(2s+3( 2520s^10+12600s^9+1750 s^8 -68600 s^7)\\ &&&-13130s^6+253630 s^5-14558s^4-557066 s^3+206601 s^2+542619 s -360360) \\ \\
+&\frac{1}{1098240}&{g_1}^2{g_2}{ g_3}& s(s+1)(2s -3)(2s-1)(2s+1)(2s+3)(2s+5)(280s^8+1120s^7-670s^6 \\ &&&-5930s^5+3047 s^4 +17284s^3 -11237s^2 -21054s+18018)\\ \\
+&\frac{1}{3294720}&{g_1}{g_2}^3&s(s+1)(2s-1)(2s+1)(2s+3) (1120 s^{10}+5600s^9-2400s^8-43200s^7\\ &&&-8814s^6+201162 s^5-60127 s^4-517124s^3+256797s^2+557766s-405405)\\ \\+&\frac {1}{274560}&{g_1}{g_3} ^2&s^2(s+1)^2(2s -3)(2s-1)(2s+1)(2s+3)(2s+5)\\ &&&(80s^6 +240 s^5-840 s^4-2080 s^3+4401 s^2+5481 s-7425)\\\\ +&\frac{1}{274560}& g_2 ^2 g_3&s(s+1)(2s-3)(2s-1)(2s+1)(2s+3)(2s+5)(80 s^8 +320 s^7\\ &&&-600 s^6-2920 s^5+4037 s^4+13314 s^3-16959 s^2-24156 s+27027)
  \end{array}$
\normalsize

\bigskip
%\newpage 

%\addcontentsline{toc}{section}{\underline{References}}
\end{document}